\documentclass[11pt,twoside]{article}

\usepackage{asp2014}

\aspSuppressVolSlug
\resetcounters

\begin{document}

\title{Astronomical research in the next decade: trends, barriers and needs in data access, management, visualization and analysis.}

\author{Cristobal~Bordiu,$^1$ Filomena~Bufano,$^1$ Eva~Sciacca,$^1$ Simone~Riggi, $^1$ Marco~Molinaro, $^1$ Giuseppe~Vizzari, $^2$ Mel~Krokos, $^3$ and Carlos~Brandt $^4$ 
\affil{$^1$Istituto Nazionale di Astrofisica (INAF), IT\\
$^2$University of Milano-Bicocca, IT\\
$^3$ University of Portsmouth, UK\\
$^4$ Jacobs University Bremen, DE\\
\email{cristobal.bordiu@inaf.it}
}}

\paperauthor{Cristobal~Bordiu}{cristobal.bordiu@inaf.it}{0000-0002-7703-0692}{INAF}{Osservatorio Astrofisico di Catania}{Catania}{IT}{95123}{Italy}
\paperauthor{Filomena~Bufano}{filomena.bufano@inaf.it}{0000-0002-3429-2481}{INAF}{Osservatorio Astrofisico di Catania}{Catania}{IT}{95123}{Italy}
\paperauthor{Eva~Sciacca}{eva.sciacca@inaf.it}{0000-0002-5574-2787}{INAF}{Osservatorio Astrofisico di Catania}{Catania}{IT}{95123}{Italy}
\paperauthor{Marco~Molinaro}{marco.molinaro@inaf.it}{0000-0001-5028-6041}{INAF}{Osservatorio Astrofisico di Trieste}{Trieste}{IT}{34131}{Italy}
\paperauthor{Simone~Riggi}{simone.riggi@inaf.it}{0000-0001-6368-8330}{INAF}{Osservatorio Astrofisico di Catania}{Catania}{IT}{95123}{Italy}
\paperauthor{Giuseppe~Vizzari}{giuseppe.vizzari@unimib.it}{0000-0002-7916-6438}{University of Milano -- Bicocca}{CSAI Research Center}{Milano}{IT}{20126}{Italy}
\paperauthor{Mel~Krokos}{mel.krokos@port.ac.uk}{0000-0001-5149-6091}{University of Portsmouth}{School of Creative Technologies}{Portsmouth}{UK}{PO12DJ}{United Kingdom}
\paperauthor{Carlos~Brandt}{c.brandt@jacobs-university.de}{0000-0001-6679-3777}{Jacobs University Bremen}{Osservatorio Astrofisico di Catania}{Bremen}{DE}{28759}{Germany}



\begin{abstract}
We report the outcomes of a survey that explores the current practices, needs and expectations of the astrophysics community, concerning four research aspects: open science practices, data access and management, data visualization, and data analysis. The survey, involving 329 professionals from several research institutions, pinpoints significant gaps in matters such as results reproducibility, availability of visual analytics tools and adoption of Machine Learning techniques for data analysis. This research is conducted in the context of the H2020 NEANIAS project.
\end{abstract}



\section{Introduction}
Astronomy is entering uncharted territory. With the LSST, SKA and other next generation observing facilities just around the corner, researchers are facing an inevitable revolution. The looming data deluge caused by these new instruments will represent an unprecedented scientific and technological challenge, by far exceeding the current data access, visualization and analysis capabilities and imposing dramatic changes in the way of working of the astrophysics community. Successfully adapting to this new reality requires a paradigm shift, built upon the adoption of Open Science practices and the development of innovative solutions. In this context, the H2020 project NEANIAS stands up as one of the facilitators of this essential transition, providing a complete portfolio of agile, fit-for-purpose astrophysics services and creating a collaborative research ecosystem through the European Open Science Cloud \citep{P2-16_adassxxx}. However, a practical materialization of such an ecosystem is only possible if the provided services respond to actual community demands. To guarantee that NEANIAS tackles specific research needs, we decided to take the pulse of the community and gather first-hand feedback from astrophysics professionals.

\section{Methodology}

Inspired by similar research works in the healthcare area \citep{lopezperez2020}, we conducted a targeted online survey on potential end-users to identify current practices, perceived barriers and future expectations in the astrophysics community concerning four principal aspects: 1) open science practices, 2) data access and management, 3) data visualization and multiwavelength astronomy, and 4) data analysis and machine learning. The survey was anonymous, as participants were only asked to specify their speciality, their expertise level and, optionally, their gender. Then, for each topic, they were asked to rate the frequency they engage in certain practices, to identify the most urgent issue to be addressed or improved from a predefined list, and to rate their agreement with a series of statements about the near future. All the questions were optional. The survey was distributed via e-mail to multiple research and academic institutions, mainly from Spain, Italy, Portugal, Greece, France, Germany and UK, and also within a private Facebook group {\it including only} professional astronomers.

\section{Results}

A total of 329 individuals completed the survey. The gender distribution of the participants was 65\% male and 31\% female (4\% chose the option "Other" or skipped the question). The most represented specialities were \textit{Galaxies and Cosmology} (32\%), followed by \textit{Stellar physics, evolution and populations} (28\%), \textit{High energies and compact objects} (16\%) and \textit{Technology and Instrumentation} (11\%). On the contrary, \textit{Solar System and the Sun} and \textit{Extrasolar Planets} communities accounted for 4.5\% and 4\% of the participants respectively, with another 4.5\% choosing "Other". Participants were mostly senior scientists with more than ten years of experience (65\%).

From the questions in the \textbf{Open Science} section, the use of open-source software and non-proprietary data from public archives appear as widespread practices (done "always" or "often") for 85\% and 70\% of the respondents, respectively, although less than 1/3 of the participants share their own code or data products in public repositories. Interestingly, there seems to be a generalized concern about result reproducibility and transparency in publications, identified by 58.9\% of the participants as the most worrisome issue, to the detriment of the use of collaborative tools (27.5\%) and accessibility to journal publications (13.6\%). Finally, the community perceives that the degree of adoption of Open Science practices will increase in the next few years, with most of the participants moderately or fully agreeing that collaborative and distributed approaches will be essential for breakthrough discoveries in the next decade (77.4\%), and that data processing and analysis will be done remotely in cloud infrastructures (70\%), mostly relying on open source tools (80.5\%).

The \textbf{Data Access and Management} section yields interesting results, as well. Despite IVOA efforts and recommendations, Virtual Observatory tools are still far from being the preferred option for data access (23\% of the respondents never use them and 51\% rarely or occasionally). However, findability and interoperability of data (raw, calibrated and high-level products) are perceived as the most urgent need by 59.3\% of the surveyees, and 73.3\% of them moderately or fully agree that enforcement of standards (such as the IVOA recommendations) will become more necessary in the future. A significant fraction of the participants deal occasionally (25.3\%), often (30.8\%) or always (6.1\%) with datasets exceeding 100 GB in size, in agreement with the majority's perception of storage as one of the major issues for astrophysics in the next decade (42.7\% moderately agree and 36\% fully agree). Last but not least, seamless integration of data archives and computing facilities is deemed as necessary to allow researchers to focus on scientific production (44.2\% moderately agree and 38\% fully agree).

The outcomes of the \textbf{Data Visualization} section reflect that more than half of the participants (53\%) work often or always with multidimensional datasets. On the contrary, the use of online visualization tools seems to be sporadic, with only 1/5 of the surveyees resorting to them on a daily basis (often or always). Surprisingly, the usage of mosaics covering large areas of the sky is residual, with most of the respondents rarely (30.2\%) or never (34.8\%) needing them. Availability of Visual Analytics tools stands out as the most immediate limitation according to 71.7\% of the respondents, over other potential issues such as the availability for pipelines for mosaicking (13.1\%) or tools for large scale data visualization (15.2\%). Looking at the future, the role of Virtual and Augmented Reality in astronomy is uncertain, with 42\% of the participants not sure about its utility and applications. Nevertheless, there is a general agreement (80\%) on the need of novel visualization techniques for better dissemination of results. On the other hand, the increasing importance of multiwavelength astronomy becomes evident, with most of the participants moderately (37\%) or strongly agreeing (32.1\%) on its crucial role to unlock breakthrough scientific discoveries in the next decade.

Finally, the responses about \textbf{Data Analysis and Machine Learning} portray a somewhat contradictory landscape. 55\% of the surveyed researchers frequently work with data from all-sky surveys, and 60.7\% investigate (often or always) statistical properties of families or groups of objects. Conversely, less than 1/5 (18.5\%) makes regular use of ML techniques, with a surprising 57.3\% rarely or never using them. These trends indicate the prevalence of traditional statistical methods over more innovative approaches. However, the community acknowledges the potential of Machine Learning: while respondents show division of opinion on whether next-generation all-sky surveys will replace targeted observations (35.2\% agree and 37.6\% disagree), 45\% of them moderately or strongly agree that many scientific discoveries will be achieved employing Machine Learning-supported processes, with little or no human intervention. In this sense, a detailed discussion of the validity of these methods, their possible biases, limitations and consequences is perceived as necessary by 88.1\% of the surveyees.
Source cross-matching in all-sky surveys constitutes one of the main barriers in Data Analysis for almost half of the participants (44.9\%), followed by the availability of Machine Learning libraries, tools and documentation tailored for astrophysics (39.7\%).
 
\begin{table}[!ht]
\caption{Research barriers per speciality (in \%). \textit{Galaxies and Cosmology (GC)}, \textit{Stellar physics (SP)}, \textit{Technology and Instrumentation (TI)}, \textit{High Energies and Compact Objects (HE)}, \textit{Solar System and the Sun (SS)} and \textit{Extrasolar Planets (EP)}}
\label{tab:barriers}
\smallskip
\begin{center}
{\small
\begin{tabular}{llllllll}  
\tableline
\noalign{\smallskip}
Research barrier & GC & SP & TI & HE & SS & EP & Total \\
\noalign{\smallskip}
\tableline
\noalign{\smallskip}
Reproducibility/transparency of results & 60.4 & 58.4 & 62.8 & 66.0 & 40.0 & 46.1 & 58.9\\
Adoption of collaborative workflows & 30.7 & 29.2 & 28.6 & 18.0 & 20.0 & 23.1 & 27.5 \\
Accessibility of journal publications & 8.9 & 12.4 & 8.6 & 16.0 & 40.0 & 30.8 & 13.6\\
\noalign{\smallskip}
\tableline 
\noalign{\smallskip}
Findability and interoperability of data & 53.9 & 59.3 & 55.9 & 72.0 & 60.0 & 53.8 & 59.3 \\
Infrastructure of data archives & 15.3 & 23.1 & 23.5 & 12.0 & 6.7 & 23.1 & 17.7\\
Availability of computational resources & 30.8 & 17.6 & 20.6 & 16.0 & 33.3 & 23.1 & 23.0 \\
\noalign{\smallskip}
\tableline 
Tools for large scale data visualization & 18.6 & 17.8 & 8.6 & 12.5 & 6.7 & 0.0 & 15.2\\
Pipelines for mosaic image generation & 17.6 & 11.1 & 5.7 & 14.6 & 6.7 & 8.3 & 13.1 \\
Availability of Visual Analytics tools & 63.7 & 71.1 & 85.7 & 72.9 & 86.7 & 91.7 & 71.7\\
\noalign{\smallskip}
\tableline 
\noalign{\smallskip}
Source cross-match in all-sky surveys & 48.5 & 45.5 & 29.0 & 54.9 & 21.4 & 30.8 & 44.9 \\
Availability of ML tools/tutorials & 38.6 & 39.8 & 38.7 & 33.3 & 57.1 & 61.5 & 39.7\\
Availability of annotated ML datasets & 12.9 & 14.8 & 32.3 & 11.8 & 21.4 & 7.7 & 15.4 \\
\noalign{\smallskip}
\tableline\
\end{tabular}
}
\end{center}
\end{table}

\section{Discussion and conclusions}

The survey presented in this work involves a representative sample of professionals, and thus provides valuable insights about the way of working of the astrophysics community, allowing us to draw some interesting conclusions. We observe a certain asymmetry concerning data and resource sharing. This, along with the positive perception of these practices and its growing importance, may indicate that the adoption of the Open Science culture is still at early stages. With respect to research barriers and issues, summarized in Table \ref{tab:barriers}, the trends hold when disaggregated per speciality. The concern about results reproducibility is particularly noteworthy. In spite of observatory policies for data release, and the increasing availability of manuscripts in public repositories such as ArXiv, significant advances in methodological transparency and sharing of processing scripts and data products are still indispensable. Similarly, data findability and interoperability emerge as two crucial aspects in the future of astrophysics research. These issues underline the need of integrating the FAIR principles (Findability, Accessibility, Interoperability and Reusability), which indeed are pillars of the NEANIAS project, into the astronomical data life cycle. Equally compelling is the wide demand for Visual Analytics tools. NEANIAS will deliver state-of-the-art solutions integrating data visualization and analysis (VLVA, \citealt{vitello2018vialactea}) and enhanced data accessibility (VLKB, \citealt{molinaro2016vialactea}). Lastly, the apparent gap in the adoption of Machine Learning outlines a unique opportunity, considering that a significant part of the community is involved in large scale statistical studies of stellar populations, galaxies and other sources. The overwhelming data volumes expected in next-generation surveys will soon render traditional approaches impractical, and therefore Machine Learning is called to become one of the spearheads of scientific discovery in the next decade. In this regard, NEANIAS Machine Learning services are specifically tailored to automate the extraction, characterization and classification of compact and extended sources \citep{riggi2019} from all-sky surveys.

\acknowledgements The research leading to these results has received funding from the European Commissions Horizon 2020 research and innovation programme under the grant agreement No. 863448 (NEANIAS).

\bibliography{O11-96}  

\begin{thebibliography}{}
\expandafter\ifx\csname natexlab\endcsname\relax\def\natexlab#1{#1}\fi
\expandafter\ifx\csname url\endcsname\relax
  \def\url#1{\texttt{#1}}\fi
\expandafter\ifx\csname urlprefix\endcsname\relax\def\urlprefix{URL }\fi
\providecommand{\eprint}[2][]{\url{#2}}

\bibitem[{Lopez-Perez et~al.(2020)Lopez-Perez, Canevari, Pecchia, Arredondo,
  Licitra, \& Fico}]{lopezperez2020}
Lopez-Perez, L., Canevari, S., Pecchia, L., Arredondo, M., Licitra, L., \&
  Fico, G. 2020, IFMBE Proceedings, 74, 174

\bibitem[{Molinaro et~al.(2016)}]{molinaro2016vialactea}
Molinaro, M., et~al. 2016, in Software and Cyberinfrastructure for Astronomy IV
  (International Society for Optics and Photonics), vol. 9913, 99130H

\bibitem[{Riggi et~al.(2019)}]{riggi2019}
Riggi, S., et~al. 2019, Publications of the Astronomical Society of Australia,
  36

\bibitem[{{Sciacca}(2021)}]{P2-16_adassxxx}
{Sciacca}, E. 2021, in ADASS XXX, edited by J.-E. {Ruiz}, \& F.~{Pierfederici}
  (San Francisco: ASP), vol. TBD of ASP Conf. Ser., 999 TBD

\bibitem[{Vitello et~al.(2018)}]{vitello2018vialactea}
Vitello, F., et~al. 2018, Publications of the Astronomical Society of the
  Pacific, 130, 084503

\end{thebibliography}


\end{document}